%% file: prl_final.tex
\documentclass[aps,prl,twocolumn,superscriptaddress,showpacs]{revtex4}

\usepackage{graphicx}
\usepackage{dcolumn}
\usepackage{bm}

\def\ap#1,#2,#3#4{           {\it Ann. Phys. (NY)\/ }{\bf #1} #2 (19#3#4)}
\def\apj#1,#2,#3#4{          {\it Astrophys. J.\/ }{\bf #1} #2 (19#3#4)}
\def\apjl#1,#2,#3#4{         {\it Astrophys. J. Lett.\/ }{\bf #1} #2 (19#3#4)}
\def\app#1,#2,#3#4{          {\it Acta Phys. Polon.\/ }{\bf #1} #2 (19#3#4)}
\def\com#1,#2,#3#4{          {\it Comm. Math. Phys.\/ }{\bf #1} #2 (19#3#4)}
\def\ib#1,#2,#3#4{           {\it ibid.\/ }{\bf #1} #2 (19#3#4)}
\def\nat#1,#2,#3#4{          {\it Nature (London)\/ }{\bf #1} #2 (19#3#4)}
\def\np#1,#2,#3#4{           {\it Nucl. Phys.\/ }{\bf B#1} #2 (19#3#4)}
\def\npps#1,#2,#3#4{         {\it Nucl. Phys. B (Proc. Suppl.)\/ }{\bf B#1}
                             #2 (19#3#4)}
\def\plb#1,#2,#3#4{          {\it Phys. Lett.\/ }{\bf B#1} #2 (19#3#4)}
\def\pla#1,#2,#3#4{          {\it Phys. Lett.\/ }{\bf A#1} #2 (19#3#4)}
\def\prd#1,#2,#3#4{          {\it Phys. Rev.\/ }{\bf D#1} #2 (19#3#4)}
\def\prep#1,#2,#3#4{         {\it Phys. Rep.\/ }{\bf #1} #2 (19#3#4)}
\def\prl#1,#2,#3#4{          {\it Phys. Rev. Lett.\/ }{\bf #1} #2 (19#3#4)}
\def\pro#1,#2,#3#4{          {\it Prog. Theor. Phys.\/ }{\bf #1} #2 (19#3#4)}
\def\rmp#1,#2,#3#4{          {\it Rev. Mod. Phys.\/ }{\bf #1} #2 (19#3#4)}
\def\sp#1,#2,#3#4{           {\it Sov. Phys.-Usp.\/ }{\bf #1} #2 (19#3#4)}

\def\zp#1,#2,#3#4{           {\it Zeit. fur Physik\/ }{\bf C#1} #2 (19#3#4)}

\def\simge{\mathrel{%
   \rlap{\raise 0.511ex \hbox{$>$}}{\lower 0.511ex \hbox{$\sim$}}}}
\def\simle{\mathrel{
   \rlap{\raise 0.511ex \hbox{$<$}}{\lower 0.511ex \hbox{$\sim$}}}}
\def\lessim{\mathrel {\vcenter {\baselineskip 0pt \kern 0pt
\hbox{$<$} \kern 0pt \hbox{$\sim$} }}}
\def\gessim{\mathrel {\vcenter {\baselineskip 0pt \kern 0pt
\hbox{$>$} \kern 0pt \hbox{$\sim$} }}}
\def\ppb{\mbox{$p\bar{p}$}}
\def\ol{\overline}

\def\sm{standard model}
\def\md{M_D}
\def\met{\mbox{$\not\!\!{E}_{T}\,$}}
\def\et{\mbox{$E_T$}}
\def\pt{\mbox{$p_T$}}

\def\Zll{\mbox{$Z\rightarrow \ell \ell$}}
\def\Znn{\mbox{$Z\rightarrow\nu\overline{\nu}$}}
\def\Zllp{\mbox{$Z(\rightarrow \ell \ell)$}}
\def\Znnp{\mbox{$Z(\rightarrow\nu\overline{\nu})$}}
\def\Wen{\mbox{$W\rightarrow e\nu$}}
\def\Wmn{\mbox{$W\rightarrow\mu\nu$}}
\def\Wtn{\mbox{$W\rightarrow\tau\nu$}}
\def\Wln{\mbox{$W\rightarrow\ell\nu$}}
\def\Wlnp{\mbox{$W(\rightarrow\ell\nu)$}}
\def\qqgG{\mbox{$q\overline{q} \rightarrow gG$}}
\def\qgqG{\mbox{$qg \rightarrow qG$}}
\def\gggG{\mbox{$gg \rightarrow gG$}}
\def\pyt{{\sc pythia}}
\def\geant3{{\sc geant3}}
\def\ptmin{\mbox{$P_T^{\rm{min}}$}}
\def \rightdownarrow
 {\kern.4em {\raise1.75ex\hbox{$\Bigl |$}$\kern-0.27em{\longrightarrow}$}}
\def \mrightdownarrow
 {\kern.4em {\raise1.25ex \hbox{$|$} \kern-0.27em{\longrightarrow}}}

\def\mMEt{\not\kern-.35em {E_T}}
\def\MEt{\hbox{$\mMEt$}}

\begin{document}
%
%


\title{Search for Large Extra Dimensions in the Production
of Jets and Missing Transverse Energy in $p\bar{p}$ Collisions at
$\sqrt{s}$ = 1.96 TeV}


%
\input{January_2006_Authors.tex}
%
%
%
%

\begin{abstract}
We present the results of a search for new physics in the 
jets plus missing transverse energy data sample collected 
from 368~pb$^{-1}$ of \ppb~ collisions at $\sqrt{\mbox{s}} 
=$1.96~TeV recorded by the Collider Detector at Fermilab.
We compare the number of events observed in the data with 
a data-based estimate of the \sm\ backgrounds 
contributing to this signature.  We observe no significant 
excess of events, and we interpret this null result in terms 
of lower limits on the fundamental Planck scale for a large 
extra dimensions scenario.
\end{abstract}

\pacs{14.80.-j, 04.50.+h, 13.85.Rm}
\maketitle


One of the simplest new physics signatures that can be explored
at a hadron collider consists of a very energetic jet and large missing 
transverse energy ($\met$).  The Tevatron offers a unique opportunity to
explore energy regimes that could yield new physics that have
not been accessible at previous colliders.  We take advantage of
this opportunity by performing a signature-based, high energy 
monojet search.  While a wide range of exotic 
physics both known and not yet imagined could yield such a signature,
the most exciting recent scenario involves Large Extra Dimensions 
(LED)~\cite{grw}. LED are an essential ingredient of proposed 
solutions to the most fundamental problems of physics including 
the hierarchy problem~\cite{add} and the observed value of the dark 
energy ~\cite{cliff}.



In LED scenarios,  gravitons or their superpartners~\cite{cliff,hugo} 
are responsible for the observed $\met$. 
In the simplest $2\rightarrow 2$ tree-level processes possible at a hadron 
collider experiment, any of the gravitational states (we denote all possible 
states with spins from 0 to 2 by $G$) can be directly produced in 
processes such as $\qqgG$, $\qgqG$, and $\gggG$, leaving the final 
state quark ($q$) or gluon ($g$) to produce a single jet~\cite{grw}. 


 
Graviton emission is within reach of the Tevatron provided 
that the ($4+n$)-dimensional Planck scale ($M_D$) is around
1~TeV~\cite{grw,add}.  This can be the case if the radii $R$ 
of the $n$ compactified extra dimensions are sufficiently 
large.  Assuming $n$ extra dimensions of the same size, the 
relationship between the 4-dimensional effective Planck scale 
$M_{Pl} \sim 10^{19}$ GeV and the fundamental Planck scale 
$M_D$ is given by the generalized Gauss theorem~\cite{add}:

\begin{equation}
M_{Pl}^2 = 8 \pi  R^n M^{2+n}_{D}.  
\label{gauss}
\end{equation}
In the absence of a significant excess of events 
relative to the background expectations, lower limits can be 
set on $M_D$, which is a fundamental parameter common to all 
LED models.

Monojet searches performed in Run~I~\cite{Run1jetmet,D0Run1} at the 
Tevatron were consistent with \sm\ (SM) expectations.  In this Letter 
we present a new search using 368 pb$^{-1}$ of data, recorded 
by the Collider Detector at Fermilab (CDF) in Run~II of the 
Tevatron.   The sensitivity of this analysis to new physics in general and
to LED scenarios in particular, is significantly 
improved by the increase in center of mass energy for Run~II 
$\ppb$ collisions (1.80~TeV to 1.96~TeV), an improved CDF 
detector, and a factor of four increase in the integrated 
luminosity over the data sample used in Run~I.

A complete description of the CDF~II detector is given in 
Ref.~\cite{tdr,svx}.  The important components used in the 
reconstruction of the events for this analysis include a 
tracking system consisting of a silicon-strip vertex detector 
surrounded by an open-cell drift chamber.  The tracking system 
is situated within a $1.4$~T solenoidal magnetic field to allow 
for the measurement of charged particle momenta transverse to 
the beamline ($p_{T}$). Outside the magnet, scintillator-based 
electromagnetic and hadronic calorimeter modules are arranged 
in projective tower geometries to reconstruct the energy and 
direction of the particle jets.   
The outermost detection system consists of planes of multi-layered 
drift chambers for detecting muons.  

We select events in the trigger system based on the presence of 
a jet with transverse energy $E_T >$100~GeV and further require 
offline that the highest $E_T$ jet in the event (the leading jet) 
has $E_T >$150~GeV.  With an offline threshold of 150~GeV, the jet 
trigger is $>$~99~$\!\%$ efficient for the events in our sample.  
Jets are reconstructed using a fixed cone algorithm with cone size 
$\Delta\mathcal{R}\equiv\sqrt{(\Delta\eta)^2 + (\Delta\phi)^2} 
= 0.7$~\cite{coordinates} over the full pseudorapidity coverage 
of the calorimeters ($|\eta| <$~3.6).  The jet $E_{T}$ are 
corrected in order to account for the effects of fragmentation, 
calorimeter non-uniformities, and energy from the rest of the 
event~\cite{jetcor}.  To increase the acceptance for events 
containing a second jet originating from quark or gluon radiation, 
we accept events that contain a second jet with $E_{T} <$60~GeV.  
However, we reject events that contain three or more reconstructed 
jets with $E_{T} >$20~GeV because these events have a much lower 
signal-to-background ratio than one-jet and two-jet events.  Since 
the \met~\cite{metdef} 
for most backgrounds is typically much lower than that 
of the signal, we require $\met >$120~GeV.  This value is set 
lower than the jet $E_T$ threshold of 150~GeV to keep the signal 
efficiency high given the $\met$ resolution.

To further reduce backgrounds from multi-jet events in which one 
or more of the jets are badly mis-measured, we require that the 
azimuth $\phi$ of the observed $\met$ is separated by more than 
0.3 radians from the $\phi$ of any second jet.  

To eliminate non-collision background events originating from 
beam halo, cosmic rays, and detector noise, we require that the 
leading jet is central ($|\eta| < 1.0$) and contained within the 
instrumented parts of the calorimeter.  Since a jet with $|\eta| 
< 1.0$ is within the fiducial tracking volume, we can search for 
tracks pointing towards the region of the calorimeter in which 
the jet is found.  For all jets within the fiducial tracking 
volume, we require at least two associated tracks whose $p_{T}$ 
add up to at least 10\% of the jet $E_{T}$.  
We also require an event vertex reconstructed from six or more 
tracks that is within 60~cm of the detector center in $z$ (the 
coordinate parallel to the colliding beams).  For the leading,
central jet we also require that the associated tracks used 
in the $p_{T}$ sum described above are consistent 
with having orginated from this event vertex.  To eliminate 
background from muons produced upstream of the detector 
that interact and look like jets in the hadronic calorimeter,
we additionally require that for each event the total 
electromagnetic energy of all jets with $E_{T} >$20~GeV is 
at least 10\% of the total $E_{T}$. 

The resulting candidate sample contains a significant number 
of events originating from SM processes which can    
produce large $\met$ in the detector.  The $\met$ associated 
with these processes can originate from either neutrinos in 
the final state (real $\met$) or other particles that pass   
into uninstrumented regions of the detector (fake $\met$).  
The largest SM background is $Z$+jets where the $Z$ boson 
subsequently decays into neutrinos ($Z\rightarrow\nu\ol{\nu}$). 
This background has the same event topology as our signal and 
is thus irreducible.  The next most significant SM background 
comes from $\Wlnp$+jets production ($\ell = e$, $\mu$, or 
$\tau$) where the lepton is unidentified. 
The contribution of this background is suppressed by rejecting 
events that contain an isolated track with $p_{T} >$~10~GeV/$c$ 
(a potential muon) or a jet with $E_{T} >$~20~GeV for which 
the electromagnetic energy fraction is above 90\% (a potential 
electron).  Track isolation is defined using the measured 
energy in the calorimeter within a $\Delta\mathcal{R} <$~0.4 
cone around the reconstructed track, after subtracting the 
measured energy in those calorimeter towers intersected by the 
track.  Tracks are defined to be isolated  if this energy is 
less than 10~$\!\%$ of the measured track $p_{T}$.  We refer 
to the $W/Z$+jets backgrounds collectively as electroweak 
backgrounds.  

The number of electroweak background events in the candidate 
sample is estimated by measuring cross sections for $\Zllp$+jets 
and $\Wlnp$+jets ($\ell = e$ or $\mu$) production from independent 
data samples collected using high $E_{T}$ single electron and 
high $p_{T}$ single muon triggers.  
We select events that contain a muon with $\pt >$20~GeV/$c$ or 
an electron with $\et >$25~GeV using standard lepton selection 
criteria~\cite{wzprl} to constuct a low-background sample of   
lepton candidates.  Starting from this sample, we select $\Wln$ 
candidates by requiring $\met >$25~GeV ($\met >$20~GeV for muon 
events) and $\Zll$ candidates by requiring a second lepton that
satifies a looser set of selection criteria~\cite{wzprl}.  The 
cosmic ray background in both candidate samples is reduced 
by rejecting events in which tracks passing through opposite 
hemispheres of the detector can be reconstructed along a common 
trajectory. 

Using the measured $\Zllp$+jets cross sections and 
the difference in $Z$ branching fractions for charged 
leptons and neutrinos, we estimate the expected number 
of $\Znnp$+jets events in our candidate sample.  A 
second, independent estimate of this background is 
obtained from the measured $\Wlnp$+jets cross sections.  
In this case we first divide the measured cross section 
by a theoretical prediction for $R_{(W/Z)}$, the ratio 
of the $W$+jets and $Z$+jets production cross sections 
and then correct the extrapolated $\Zllp$+jets cross 
section for the $Z$ branching fraction to neutrinos.  
A more precise prediction for the expected background 
is obtained by combining the estimates.

\begin{figure}[tb]
\begin{center}
\includegraphics[width=3.2in]{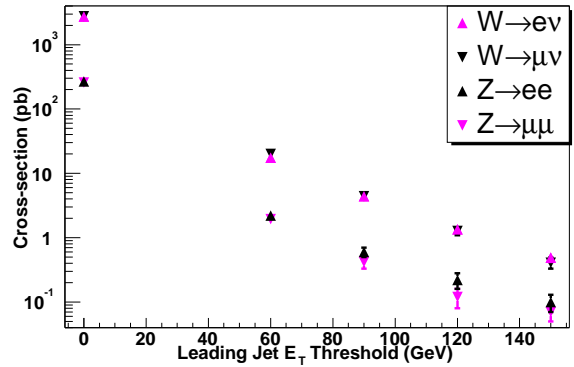} 
\caption{$W$ and $Z$ Boson cross sections as a function of 
leading jet $E_T$.}
\end{center}
\label{WZxsecFig}
\end{figure}

As a consistency check of the event selection,
we measure inclusive 
cross sections for $W/Z$ production and compare the results 
with published Run~II measurements~\cite{wzprl}.  In addition,
we measure $W/Z$+jet production cross sections where the jet 
criteria are identical to those used in the final selection of 
our jet plus $\met$ candidate sample, except that
we vary the leading jet $E_T$ cut for these measurements,
using thresholds of 60, 90, 120, and 150~GeV. 
The larger statistics available in the samples obtained using 
the lower jet $E_{T}$ thresholds allow for statistically 
significant comparisons between independent measurements in 
the electron and muon channels.  The observed agreement 
provides further validation of the cross section measurements 
made for the 150 GeV leading jet threshold that are used to 
estimate the electroweak backgrounds in our final candidate 
sample.

The acceptance values used in the $W/Z$+jets cross section 
measurements are obtained from simulated \pyt~\cite{pythia} 
event samples using a full detector simulation based on 
\geant3~\cite{geant} and corrected to account for measured 
differences in lepton selection criteria observed in data 
and simulation.  The measured cross sections are shown in 
Fig.~1.  We combine cross section measurements from the 
electron and muon samples using the default cut of 150~GeV 
on the $E_{T}$ of the leading jet and obtain
$\sigma(\Wlnp+{\rm{jets}})~=~0.46~\pm~0.05~\rm{pb}$ and
$\sigma(\Zllp+{\rm{jets}})~=~0.08~\pm~0.02~\rm{pb}.$  

A next-to-leading order calculation based on the {\sc mcfm} 
generator~\cite{mcfm} was used to determine the value 
of $R_{(W/Z)}$.  
At $\ptmin$=150 GeV, the calculated value of 
$R_{(W/Z)}$ is 8.15~$\pm$~0.40.  Based on the 
combined $\Zllp$+jets and $\Wlnp$+jets cross 
section measurements, we estimate 177~$\pm$~44 and 
125~$\pm$~15 background events from $\Znnp$+jets 
in our jet plus $\met$ candidate sample.  The two 
independent results are combined to obtain a final 
estimate of 130~$\pm$~14 events.  Estimates for the 
contributions of the other electroweak backgrounds 
to the candidate sample are also obtained from the 
measured $W/Z$+jets cross sections.  We extract, 
for example, the number of $\Wlnp$+jets background 
events from the measured $\Wlnp$+jets cross section 
based on the percentage of simulated $\Wlnp$+jets
events which pass our selection criteria.
Since the same 
measured cross sections are used to estimate the 
contributions of each electroweak background shown 
in Table~\ref{all_bgrs}, the uncertainties on these 
background predictions are fully correlated. 


As described previously, events originating from 
QCD multi-jet production processes can enter our 
sample when the mis-measurement of one or more jets 
creates large $\met$ in the detector.  The dominant 
topology is two-jet events where the second jet is 
not found by the jet finding algorithm.  To estimate 
the background contribution of such events, we study 
dijet events in data for which the observed $\met$ 
points in the direction of the less energetic jet.
We perform a linear extrapolation of the $E_{T}$
distribution for this less energetic jet into the 
region where the $E_T$ drops below our threshold 
(20~GeV) for defining jets.  Monte Carlo studies 
indicate that an additional relative contribution 
of approximately 15\% from three-jet events should 
be added to the number extracted from this method, 
resulting in an estimate of $15 \pm 10$ background
events from QCD multi-jet production in our final 
candidate sample.


Using timing information from the hadronic calorimeter 
we estimate the non-collision background, from sources 
such as cosmic rays, to be 4~$\pm$~4 events.  

\begin{figure}[tb]
\begin{center}
\includegraphics[width=3.2in]{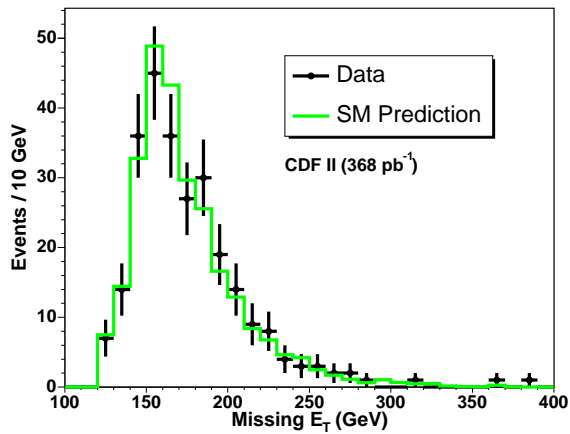} 
\caption{Comparison of the event $\met$ distribution 
         for the 263 events in our candidate sample 
         with the predicted SM distribution.}
\end{center}
\label{METdist}
\end{figure}

A summary of estimated background contributions to 
the candidate sample is shown in Table~\ref{all_bgrs}.  
We predict a total background of 265~$\pm$~30 events,
observe 263 events in the data, and therefore 
conclude that no excess is present in the data. 
Figure~2 shows a comparison of the $\met$ distribution 
for the 263 events in our candidate sample with the 
expected distribution from the SM backgrounds.  We 
note that shape of the distribution for signal events would 
not look significantly different than that shown here 
for the SM processes.
  
\begin{table}
\begin{center}
\begin{tabular}{|l|c|} \hline
{Background}     & {Events}       \\\hline\hline
{$\Znn$}         & {130 $\pm$ 14} \\\hline
{$\Wtn$}         & {60  $\pm$ 7}  \\\hline
{$\Wmn$}         & {36  $\pm$ 4}  \\\hline
{$\Wen$}         & {17  $\pm$ 2}  \\\hline
{$\Zll$}         & { 3  $\pm$ 1}  \\\hline
{QCD multi-jet}  & {15  $\pm$ 10} \\\hline\hline
{Non-collision}  & {4  $\pm$ 4}   \\\hline\hline
{Total expected} & {265 $\pm$ 30} \\\hline
{Data observed}  & {263}          \\\hline
\end{tabular}
\end{center}
\caption{Summary of estimated background 
         contributions and number of events 
         observed for the candidate sample.}
\label{all_bgrs}
\end{table}


\begin{figure}[tb]
\begin{center}
\includegraphics[width=3.2in]{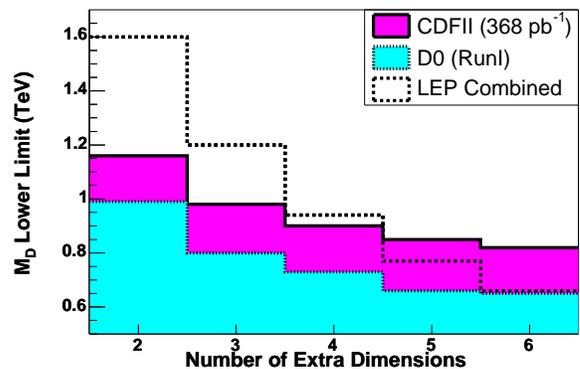} 
\caption{Comparison of 95\% CL lower limits on $M_D$ 
         based on the Run~II CDF and Run~I D\O\ 
         results with LEP combined results.}
\end{center}
\label{LimitPlot}
\end{figure}

We set lower limits on the ($4+n$)-dimensional Planck scale 
using these results. We use \pyt\ in conjuction with the 
full detector simulation 
to generate samples of simulated graviton production, 
based on leading-order production 
cross sections calculated in Ref.~\cite{grw}.  We simulate 
the signal processes for numbers of extra dimensions   
between 2 and 6 and for a set of different $\md$ values.
The cumulative signal acceptance for our selection criteria 
ranges from 9.9 $\pm$ 1.3~$\!\%$ to 12.6 $\pm$ 1.7~$\!\%$ 
as a function of $n$. The acceptances are found to have no 
significant dependence on $\md$.
Contributions to the uncertainty on the acceptance include 
the choice of parton distribution functions (5.9\% relative 
uncertainty), possible differences in the jet energy 
scale~\cite{jetcor} between data and simulation (8\%),
the models for initial and final state radiation in the
Monte Carlo simulation (5.2\%), and the uncertainty on the integrated 
luminosity of the sample (6\%).  We scale the generated 
cross sections by a K-factor (the expected ratio of cross 
sections as calculated at next-to leading order and leading 
order) of 1.3~\cite{kfact}.

We obtain the upper limit on the number of signal events
in our candidate sample using a Bayesian approach with 
a flat prior for the number of signal events and gamma 
distributions for the priors for both the acceptance 
and the number of background events~\cite{statref}.
Based on 263 observed events, a SM expectation of 
265~$\pm$~30 events, and a combined uncertainty of 
13.2~$\!\%$ on the signal acceptance, we obtain an 
upper limit of 67 signal events at 95\% C.L.,
corresponding to a cross-section times acceptance 
of (67/368 $pb^{-1}$)=0.18~pb.  We set limits on the value 
of $\md$ for different values of $n$ based on the 
maximum possible number of observed signal events.  
The lower limits on $\md$ for $n =$~2--6 are shown 
in Table~II for $K =$~1.3.  
\begin{table}
\begin{center}
\begin{tabular}{|l|c|c|}\hline
      {n} & {$M_D$(TeV)} &{$R$(mm)}               \\ \hline\hline
      2  & $>~1.18$     & $<~3.5 \times 10^{-1}$  \\
      3  & $>~0.99$     & $<~3.6 \times 10^{-6}$  \\
      4  & $>~0.91$     & $<~1.1 \times 10^{-8}$  \\
      5  & $>~0.86$     & $<~3.5 \times 10^{-10}$ \\
      6  & $>~0.83$     & $<~3.4 \times 10^{-11}$ \\ \hline
\end{tabular}
\end{center} 
\caption{95\% C.L. limits on $\md$ for $n$=2--6, 
         with the corresponding limit on $R$, 
         assuming toroidal compactification as 
         described in~\cite{grw}.}
\label{lim_mdr}
\end{table}
Assuming compactification on a torus, these limits 
on $\md$ can easily be related to limits on the 
radius of the extra dimensions by Eq.~(\ref{gauss}).


A comparison of the limits on $M_D$ from D\O\ 
and CDF with LEP combined results~\cite{lepresults} is 
shown graphically in Fig.~3.  This measurement places 
the most stringent limits from the Tevatron and in the 
case of $n =$~5 and $n =$~6 the world's best limits on 
$M_D$.  For the case of two extra dimensions, the upper 
limit of 0.35~mm on the size of the extra dimensions 
based on the lower limit for $M_D$ (1.18~TeV) can be 
compared to the limit of 0.13~mm from a direct probe 
of gravity at short distances~\cite{EotWash}.  Such 
experiments have no sensitivity for higher values of 
$n$ but have the best limits for $n$=2.



To summarize, we have performed a search for direct 
graviton production in a sample of events containing 
one high $E_{T}$ jet and large $\met$ obtained from 
368~pb$^{-1}$ of CDF Run~II data.  The number of 
events observed in the data is consistent with SM
expectations, and hence we set lower limits at the 
95\% C.L. on the fundamental Planck scale $\md$
for numbers of extra dimensions between 2 and 6. 

We thank the Fermilab staff and the technical staffs 
of the participating institutions for their vital contributions. 
This work was supported by the U.S. Department of Energy and 
National Science Foundation; the Italian Istituto Nazionale di 
Fisica Nucleare; the Ministry of Education, Culture, Sports, 
Science and Technology of Japan; the Natural Sciences and 
Engineering Research Council of Canada; the National Science 
Council of the Republic of China; the Swiss National Science 
Foundation; the A.P. Sloan Foundation; the Bundesministerium 
f\"ur Bildung und Forschung, Germany; the Korean Science and 
Engineering Foundation and the Korean Research Foundation; the 
Particle Physics and Astronomy Research Council and the Royal 
Society, UK; the Russian Foundation for Basic Research; the 
Comisi\'on Interministerial de Ciencia y Tecnolog\'{\i}a, Spain; 
in part by the European Community's Human Potential Programme 
under contract HPRN-CT-2002-00292; and the Academy of Finland. 


\end{document}

%% file: January_2006_Authors.tex
\affiliation{Institute of Physics, Academia Sinica, Taipei, Taiwan 11529, Republic of China} 
\affiliation{Argonne National Laboratory, Argonne, Illinois 60439} 
\affiliation{Institut de Fisica d'Altes Energies, Universitat Autonoma de Barcelona, E-08193, Bellaterra (Barcelona), Spain} 
\affiliation{Baylor University, Waco, Texas  76798} 
\affiliation{Istituto Nazionale di Fisica Nucleare, University of Bologna, I-40127 Bologna, Italy} 
\affiliation{Brandeis University, Waltham, Massachusetts 02254} 
\affiliation{University of California, Davis, Davis, California  95616} 
\affiliation{University of California, Los Angeles, Los Angeles, California  90024} 
\affiliation{University of California, San Diego, La Jolla, California  92093} 
\affiliation{University of California, Santa Barbara, Santa Barbara, California 93106} 
\affiliation{Instituto de Fisica de Cantabria, CSIC-University of Cantabria, 39005 Santander, Spain} 
\affiliation{Carnegie Mellon University, Pittsburgh, PA  15213} 
\affiliation{Enrico Fermi Institute, University of Chicago, Chicago, Illinois 60637} 
\affiliation{Joint Institute for Nuclear Research, RU-141980 Dubna, Russia} 
\affiliation{Duke University, Durham, North Carolina  27708} 
\affiliation{Fermi National Accelerator Laboratory, Batavia, Illinois 60510} 
\affiliation{University of Florida, Gainesville, Florida  32611} 
\affiliation{Laboratori Nazionali di Frascati, Istituto Nazionale di Fisica Nucleare, I-00044 Frascati, Italy} 
\affiliation{University of Geneva, CH-1211 Geneva 4, Switzerland} 
\affiliation{Glasgow University, Glasgow G12 8QQ, United Kingdom} 
\affiliation{Harvard University, Cambridge, Massachusetts 02138} 
\affiliation{Division of High Energy Physics, Department of Physics, University of Helsinki and Helsinki Institute of Physics, FIN-00014, Helsinki, Finland} 
\affiliation{University of Illinois, Urbana, Illinois 61801} 
\affiliation{The Johns Hopkins University, Baltimore, Maryland 21218} 
\affiliation{Institut f\"{u}r Experimentelle Kernphysik, Universit\"{a}t Karlsruhe, 76128 Karlsruhe, Germany} 
\affiliation{High Energy Accelerator Research Organization (KEK), Tsukuba, Ibaraki 305, Japan} 
\affiliation{Center for High Energy Physics: Kyungpook National University, Taegu 702-701, Korea; Seoul National University, Seoul 151-742, Korea; and SungKyunKwan University, Suwon 440-746, Korea} 
\affiliation{Ernest Orlando Lawrence Berkeley National Laboratory, Berkeley, California 94720} 
\affiliation{University of Liverpool, Liverpool L69 7ZE, United Kingdom} 
\affiliation{University College London, London WC1E 6BT, United Kingdom} 
\affiliation{Centro de Investigaciones Energeticas Medioambientales y Tecnologicas, E-28040 Madrid, Spain} 
\affiliation{Massachusetts Institute of Technology, Cambridge, Massachusetts  02139} 
\affiliation{Institute of Particle Physics: McGill University, Montr\'{e}al, Canada H3A~2T8; and University of Toronto, Toronto, Canada M5S~1A7} 
\affiliation{University of Michigan, Ann Arbor, Michigan 48109} 
\affiliation{Michigan State University, East Lansing, Michigan  48824} 
\affiliation{Institution for Theoretical and Experimental Physics, ITEP, Moscow 117259, Russia} 
\affiliation{University of New Mexico, Albuquerque, New Mexico 87131} 
\affiliation{Northwestern University, Evanston, Illinois  60208} 
\affiliation{The Ohio State University, Columbus, Ohio  43210} 
\affiliation{Okayama University, Okayama 700-8530, Japan} 
\affiliation{Osaka City University, Osaka 588, Japan} 
\affiliation{University of Oxford, Oxford OX1 3RH, United Kingdom} 
\affiliation{University of Padova, Istituto Nazionale di Fisica Nucleare, Sezione di Padova-Trento, I-35131 Padova, Italy} 
\affiliation{LPNHE, Universite Pierre et Marie Curie/IN2P3-CNRS, UMR7585, Paris, F-75252 France} 
\affiliation{University of Pennsylvania, Philadelphia, Pennsylvania 19104} 
\affiliation{Istituto Nazionale di Fisica Nucleare Pisa, Universities of Pisa, Siena and Scuola Normale Superiore, I-56127 Pisa, Italy} 
\affiliation{University of Pittsburgh, Pittsburgh, Pennsylvania 15260} 
\affiliation{Purdue University, West Lafayette, Indiana 47907} 
\affiliation{University of Rochester, Rochester, New York 14627} 
\affiliation{The Rockefeller University, New York, New York 10021} 
\affiliation{Istituto Nazionale di Fisica Nucleare, Sezione di Roma 1, University of Rome ``La Sapienza," I-00185 Roma, Italy} 
\affiliation{Rutgers University, Piscataway, New Jersey 08855} 
\affiliation{Texas A\&M University, College Station, Texas 77843} 
\affiliation{Istituto Nazionale di Fisica Nucleare, University of Trieste/\ Udine, Italy} 
\affiliation{University of Tsukuba, Tsukuba, Ibaraki 305, Japan} 
\affiliation{Tufts University, Medford, Massachusetts 02155} 
\affiliation{Waseda University, Tokyo 169, Japan} 
\affiliation{Wayne State University, Detroit, Michigan  48201} 
\affiliation{University of Wisconsin, Madison, Wisconsin 53706} 
\affiliation{Yale University, New Haven, Connecticut 06520} 
\author{A.~Abulencia}
\affiliation{University of Illinois, Urbana, Illinois 61801}
\author{D.~Acosta}
\affiliation{University of Florida, Gainesville, Florida  32611}
\author{J.~Adelman}
\affiliation{Enrico Fermi Institute, University of Chicago, Chicago, Illinois 60637}
\author{T.~Affolder}
\affiliation{University of California, Santa Barbara, Santa Barbara, California 93106}
\author{T.~Akimoto}
\affiliation{University of Tsukuba, Tsukuba, Ibaraki 305, Japan}
\author{M.G.~Albrow}
\affiliation{Fermi National Accelerator Laboratory, Batavia, Illinois 60510}
\author{D.~Ambrose}
\affiliation{Fermi National Accelerator Laboratory, Batavia, Illinois 60510}
\author{S.~Amerio}
\affiliation{University of Padova, Istituto Nazionale di Fisica Nucleare, Sezione di Padova-Trento, I-35131 Padova, Italy}
\author{D.~Amidei}
\affiliation{University of Michigan, Ann Arbor, Michigan 48109}
\author{A.~Anastassov}
\affiliation{Rutgers University, Piscataway, New Jersey 08855}
% ...
\collaboration{CDF Collaboration}
\noaffiliation

%% file: prl_final.bbl
\begin{thebibliography}{20}
\bibitem{coordinates} We use a coordinate system where $\theta$\ 
is the polar angle to the proton beam, $\phi$\ is the azimuthal 
angle about this beam axis, and
$\eta$ is the pseudorapidity defined as $-\ln\tan(\theta/2)$.
\bibitem{metdef}
Missing transverse energy, \MEt, is defined as the magnitude of 
$-\sum_i E_T^i \hat{n}_i$, where $\hat{n}_i$\ is a unit vector 
in the azimuthal plane that points from the beamline to the $i$th 
calorimeter tower.
\end{thebibliography}
